# Studying Ethnic Endogamy in Croatia with a Suitable Indicator of Homophily

Anna Naszodi

In this paper, we study the patterns of ethnic endogamy in Croatia in relation to six ethnic groups between 1970 and 2015. We find that, over the 45-year period analyzed, the segmentation of the Croatian marriage market was weaker between Czechs and non-Czechs, Hungarians and non-Hungarians, Italians and non-Italians, and Slovaks and non-Slovaks than between Serbs and non-Serbs or Croats and non-Croats. This finding is substantiated by survey evidence revealing similar patterns on relative social distances between different ethnic groups in Croatia and Serbia. From a methodological perspective, we show that a plausible ranking of the degree of segmentation of the Croatian marriage market along ethnic lines can be obtained only when marital sorting is characterized by a carefully selected indicator. While a recently reinvented indicator captures sensible patterns of ethnic endogamy, the commonly applied odds-ratio fails to produce results consistent with survey evidence.

AI generated video summary of the paper: https://idil.li/wp-content/uploads/2026/04/A_Tale_of_Two_Numbers2.mp4

# 1. INTRODUCTION

Patterns of assortative mating along ethnicity is often hypothesized to reflect social distance between different ethnic groups, and also the social integration within a country (Gordon 1964, Kalmijn 1998, Lendák-Kabók 2024, Song 2009). Building on this insight, Mrdjen and Bahnik (2018) study the segmentation of Croatian society along several ethnic divides by quantifying ethnic endogamy in Croatia from 1970 to 2015 using couple-level data and the odds-ratio indicator.

This indicator, together with the odds-ratio-based IPF-algorithm, has been a popular tool for analyzing marital sorting, however, without sound reasons. As discussed in Naszodi (2025), researchers began to look for alternative indicators and methods,[1] because:

(i) the IPF produced implausible patterns of sorting when applied to data on American couples of some generations for which survey data on self-declared marital preferences provided direct evidence on the patterns under study;

(ii) some sorting patterns identified by the IPF were extremely sensitive to the chosen number of trait categories distinguished, whereas alternative methods were robust to the same choice;

(iii) the IPF fails to work in some of those cases, where no couple of a certain type is observed;

(iv) the IPF may generate counterfactuals that cannot obtain.

(v) Finally, a careful reading of the paper, Stephan and Deming (1940), written by the inventors of the IPF, shows that their algorithm was developed for a purpose different from identifying the degree of sorting through a counterfactual decomposition.

[1] See Eika, Mogstad, and Zafar (2019), Liu and Lu (2006), Macdonald (2023), and Naszodi and Mendonca (2023).

In this paper we characterize the degree of ethnic sorting in Croatia using the same data as Mrdjen and Bahnik (2018). However, *we apply a sorting indicator that is an alternative to the odds-ratio*. In particular, we use an ordinal measure of sorting, first proposed by James S. Coleman (1958). He illustrated its use by measuring the similarity of paired children. The indicator was later reinvented by Liu and Lu (2006), who illustrated its application by analyzing the degree of marital sorting in the United States. Moreover, they presented several arguments in favor of their indicator.

The *main contribution* of this paper, relative to the existing literature, is that we present new empirical findings. These also corroborate the view that the Coleman–Liu–Lu-indicator is a more suitable measure of the degree of sorting than the once-popular odds-ratio. In particular, we show that the former indicator yields more consistent patterns of ethnic sorting with some survey evidence on the relative social distances of ethnic groups. In addition, we argue that the patterns are also more consistent with our prior views formed on the basis of knowledge about the recent Croatian history.

We depart from Mrdjen and Bahnik (2018) not only by measuring homophily with an alternative indicator to the odds-ratio, but also by focusing on *fewer ethnic groups*. Their comprehensive analysis covered the marital choices of 10 groups: the Albanians, Bosniaks, Croats, Czechs, Hungarians, Italians, Roma, Serbs, Slovaks, and those who declared themselves Yugoslavs. Our analysis focuses on a restricted set of ethnic groups that meet two criteria. First, their demographic data are reliable; second, the two competing indicators – the odds-ratio and the Coleman–Liu–Lu-indicator – do not assign the same rank to their ethnic divides.

The first criterion is not met by the Bosniaks and Yugoslavs, because – as it is noted by Mrdjen and Bahnik (2018) – before 2011 Bosniaks used to declare themselves as Muslims. Similarly, there was a large variation over time in the proportion of those, who reported to have Yugoslavian identity (Kukić 2023).

The second criterion is not met by Albanians, since both the odds-ratio and the Coleman– Liu–Lu-indicator identify them to be one of the closest minority groups in Croatia. Also, by applying either of the two competing indicators, one finds the Roma population to be similarly closed as the Albanians. The extreme degree of closedness of both the Albanian and Roma groups is substantiated by the survey results of Bašić, Todosijević, Markovi´c, and Zafirović (2020) and Strabac and Valenta (2016). After applying the two criteria, we are left with six groups: Croats, Czechs, Hungarians, Italians, Serbs, and Slovaks.

We formulate now our strong *prior view on the relative social distance* of each of the six groups from the rest of society. Our knowledge of Croatian history suggests that social distance tends to be smaller for groups that were not directly involved in the Yugoslav wars of the 1990s and the related ethnic cleansing. Among the six groups, this applies to Czechs, Hungarians, Italians, and Slovaks. Accordingly, we expect ethnic endogamy to be weaker for these four minority groups than for Serbs and Croats, provided the strength of endogamy is measured with a suitable indicator.

It is important to emphasize that the patterns we expect to obtain are not intended to describe the experiences of every individual or couple. There are well-known deviations from these average tendencies, since individual partner-choices often diverge from group-level patterns. This heterogeneity implies that social distance is meaningfully defined only at the group level. Moreover, social distances are not immutable: as group compositions change over time, the distances between groups can also change. This feature suggests that the gaps examined in this paper may gradually diminish, fostering a more integrated social fabric across all ethnic groups in Croatia.

As we will see, we find the segmentation of the marriage market to be highly sensitive to the indicator chosen. In particular, if we use the *Coleman–Liu–Lu-indicator*, we find the segmentation of the marriage market to be in line with our prior view: it is weaker between Czechs and non-Czechs, Hungarians and non-Hungarians, Italians and non-Italians, and Slovaks and non-Slovaks than between Serbs and non-Serbs or Croats and non-Croats. However, if we apply the *odds-ratio*, we find the relative degrees of segmentation to be difficult to explain. In particular, similar to Mrdjen and Bahnik (2018), we get with the odds-ratio that ethnic endogamy is the weakest in relation to Croats and non-Croats, while being the second weakest for Serbs and non-Serbs in the restricted set of six ethnic groups analyzed.

The difference in rankings of the ethnic divides is consistent with the literature that recognizes the following point. The sorting or homophily indicators typically used differ primarily in how they control for the structural availability of potential partners by the assorted trait analyzed (Rosenfeld 2008). Here, we add two points related to the first one: when structural availability by ethnicity varies as much as it does across Croatian groups (see Table 1), it is not surprising that the competing indicators produce substantially different assessments of ethnic divides. On the positive side, we add that we can use the diverse set of assessments to select the suitable measure of homophily.

The rest of the paper is *structured* as follows. Section 2 describes how the two competing indicators characterize one of the determinants of the equilibrium in the marriage market. Section 3 introduces the data used in the empirical analysis. Section 4 presents two rankings of ethnic endogamy in relation to the six ethnic groups we focus on: one is obtained with the odds-ratio, while the other is obtained with the Coleman–Liu–Lu-indicator. Finally, Section 5 concludes the paper.

## 2. CHARACTERIZING ETHNIC ENDOGAMY

It has been long recognized that raw proportions of inter-ethnic couples reflect both the *structural availability* shaped by relative size of the ethnic groups and also a *directly unobserved factor*. The directly unobserved factor is of interest to those researchers, who would like to study the ethnic divide in a society by analyzing data on couples.

The literature is not consistent with the terminology used for the directly unobserved factor. All the following terms are in use: “degree of assortative mating,” “aggregate preferences for homogamy,” “degree of segmentation of the marriage market,” “strength of social barriers to intermarry,” “width of the social gap between different ethnic groups,” “degree of inclination to marry within one’s own ethnic group,” “strength of forces attracting people with the same traits,” “strength of repulsive forces between people with different traits,” and “intensity of homophily at the aggregate level.”

Despite the diversity of the terminologies used, the above factors are empirically identical: conceptually, there is only one way to compute them. In particular, their empirical identification requires to control for the structural availability.

Table 1 about here

Let us see how “controlling for the structural availability” is operationalized in the assortative mating literature in the simplest setup, where everybody is assumed to get married; and people sort into couples along a single trait, ethnicity; and they distinguish between only two groups: their own ethnic group and the rest of the society.

In this set up, the joint ethnicity distribution of couples can be represented by the following matrix

$$K = \begin{bmatrix} N_{m,m} & N_{m,x} \\ N_{x,m} & N_{x,x} \end{bmatrix}$$

Its element $N_{h,w}$ is the number of *h, w*-type marriages, where *h* denotes the husbands’ type and *w* denotes the wives’ type. Both can be from a given ethnic group (*m*-type), or outside the ethnic group (*x*-type). We denote the number of all couples by $N$ $(= N_{m,m} + N_{m,x} + N_{x,m} + N_{x,x})$.

We will use two competing indicators for ethnic endogamy, the odds-ratio and the Coleman–Liu–Lu-indicator. The odds-ratio is calculated as

$$\mathrm{OR}(K) = N_{m,m} \times N_{x,x} / (N_{m,x} \times N_{x,m}) \ . \tag{1}$$

Those researchers, who believe that the odds-ratio is the suitable measure of sorting, also believe that this indicator is immune to the relevant variations in “structural availability”, i.e., to the empirically observed variations in the row sums and column sums of *K*. Whereas those researchers, who believe that the odds-ratio has better alternatives, also believe that the alternative indicators have the prominent feature of being invariant to the relevant variations in availability.

One of the alternatives to the odds-ratio is the Coleman–Liu–Lu-indicator. The formula of the Coleman–Liu–Lu-indicator is this:

$$\mathrm{CLL}(K) = \begin{cases} [N_{x,x} - \mathrm{int}(R)] \,/\, [\min(N_{m,x} + N_{x,x}, N_{x,m} + N_{x,x}) - \mathrm{int}(R)] & \text{if } N_{x,x} \geq R \\ [N_{x,x} - \mathrm{ceil}(R)] \,/\, [\mathrm{ceil}(R) - \max(0, N_{x,m} + N_{x,x} - N_{m,m} - N_{x,m})] & \text{if } N_{x,x} < R \end{cases} \tag{2}$$

In this formula $R = (N_{x,m} + N_{x,x})(N_{m,x} + N_{x,x})/N$ is the expected number of couples where neither the wife, nor the husband is in the ethnic group *m* and the matching is random. Furthermore, int($R$) is the nearest integer to $R$ that is not higher than $R$, and ceil($R$) is the nearest integer to $R$ that is not lower than $R$.

The Coleman–Liu–Lu-indicator is bounded, as being $-1 \leq \mathrm{CLL}(K) \leq 1$. Apparently, unlike the odds-ratio, the Coleman–Liu–Lu-indicator can take negative values. However, this is just a minor difference between the two indicators. The major difference between them, is that – as we will see – some joint distributions are ranked differently by them.

We can interpret the Coleman–Liu–Lu-indicator as follows: it is the signed and normalized ordinal value assigned to the realized aggregate matching outcome *K* that is based on three benchmark hypothetical matching outcomes. One of the benchmarks is where individuals are *randomly matched*. The Coleman–Liu–Lu-indicator assigns the value 0 to the expected value of this benchmark, i.e., CLL($K_{\text{random}}$) = 0, where

$$K_{random} = \begin{bmatrix} (N_{m,m} + N_{x,m}) \times (N_{m,m} + N_{m,x})/N & (N_{m,m} + N_{m,x}) \times (N_{m,x} + N_{x,x})/N \\ (N_{m,m} + N_{x,m}) \times (N_{x,x} + N_{m,x})/N & (N_{x,m} + N_{x,x}) \times (N_{m,x} + N_{x,x})/N \end{bmatrix}$$

Another benchmark is the *maximally positive sorting*, where one can marry out of his or her ethnic group only if nobody is available in this ethnic group with opposite sex. The Coleman–Liu–Lu-indicator assigns the value +1 to this benchmark, i.e., CLL($K_{\text{max pos}}$) = +1. The third benchmark is the *maximally negative sorting*, where one can marry somebody from his or her ethnic group only if nobody is available outside this group with opposite sex. The Coleman–Liu–Lu-indicator assigns the value −1 to this benchmark, i.e., CLL($K_{\text{max neg}}$) = −1.

The construction of the Coleman–Liu–Lu-indicator exploits the following feature: when the assortative trait is dichotomous, all feasible aggregate matching outcomes with any given pair of trait distributions (i.e., the pair of trait distributions of marriageable men and women) is ranked uniquely by the number of *x,x*-type couples. In particular, the observed number of *x,x*-type couples in $K$ within the interval $[N_{x,x} - \min(N_{m,m}, N_{x,x}), N_{x,x} + \min(N_{x,m}, N_{m,x})]$ defines the ranking of $K$ within the set of feasible joint distributions.

The same ranking principle also defines the ordinal Coleman–Liu–Lu-indicator. To illustrate this point, let us assume that the observed joint distribution of couples is given by

$$K = \begin{bmatrix} 5 & 1 \\ 3 & 3 \end{bmatrix}$$

Under the same marginal distributions (or row sums and column sums) as those of $K$, the minimum and maximum number of *x,x*-type couples is 0 and 4, respectively. Accordingly, the feasible matching outcomes are $\begin{bmatrix} 2 & 4 \\ 6 & 0 \end{bmatrix}, \begin{bmatrix} 3 & 3 \\ 5 & 1 \end{bmatrix}, \begin{bmatrix} 4 & 2 \\ 4 & 2 \end{bmatrix}, \begin{bmatrix} 5 & 1 \\ 3 & 3 \end{bmatrix}, \begin{bmatrix} 6 & 0 \\ 2 & 4 \end{bmatrix}$. The ordinal CLL values are defined by the unique ranking: the CLL values assigned to these feasible matches are -1, -1/2, 0, 1/2, 1, respectively.

In case ethnicity is the assorted trait analyzed, the empirically relevant matching outcomes are those that result from non-negative sorting, i.e., where $N_{x,x} \geq R$. Under this assumption, the formula of the Coleman–Liu–Lu-indicator of (2) simplifies to

$$\text{SCLL}(K) = [N_{x,x} - \text{int}(R)]/[\min(N_{m,x} + N_{x,x}, N_{x,m} + N_{x,x}) - \text{int}(R)] \quad (3)$$

where SCLL denotes the Simplified Coleman–Liu–Lu-indicator. It can take only non-negative values.

Next, we will compare the empirical performance of the two competing indicators defined by Equations (1) and (3).

# 3. Data

We use the same data as Mrdjen and Bahnik (2018), which are shown by Tables 2 and 3.

Table 2 about here

Table 3 about here

Although Mrdjen and Bahnik (2018) do not report the joint ethnic distributions of couples, we can compute them from the statistics in Tables 2 and 3. We obtain the following joint distributions normed by the total number of couples:

$$K^{Alb.} = \begin{bmatrix} 0.0015 & 0.0015 \\ 0.0007 & 0.9963 \end{bmatrix}$$

$$K^{Bosn.} = \begin{bmatrix} 0.0037 & 0.0051 \\ 0.0033 & 0.9879 \end{bmatrix}$$

$$K^{Croats} = \begin{bmatrix} 0.7765 & 0.0639 \\ 0.0540 & 0.1057 \end{bmatrix}$$

$$K^{Czechs} = \begin{bmatrix} 0.0009 & 0.0019 \\ 0.0023 & 0.9949 \end{bmatrix}$$

$$K^{Hung.} = \begin{bmatrix} 0.0016 & 0.0028 \\ 0.0035 & 0.9920 \end{bmatrix}$$

$$K^{Ital.} = \begin{bmatrix} 0.0004 & 0.0023 \\ 0.0018 & 0.9955 \end{bmatrix}$$

$$K^{Roma} = \begin{bmatrix} 0.0014 & 0.0003 \\ 0.0003 & 0.9980 \end{bmatrix}$$

$$K^{Serbs} = \begin{bmatrix} 0.0642 & 0.0334 \\ 0.0318 & 0.8706 \end{bmatrix}$$

$$K^{Slov.} = \begin{bmatrix} 0.0002 & 0.0006 \\ 0.0007 & 0.9985 \end{bmatrix}$$

$$K^{Yugo.} = \begin{bmatrix} 0.0308 & 0.0126 \\ 0.0151 & 0.9415 \end{bmatrix}$$

# 4. Empirical findings

Once the joint ethnic distributions are known, we can compute their (Simplified) Coleman– Liu–Lu-indicators and compare them to the corresponding log odds-ratios. They are reported by Table 4.

Table 4 about here

The first thing to note when interpreting Table 4 is that *Roma are identified as the closest ethnic group* both by the Coleman–Liu–Lu-indicator and the odds-ratio. The *Albanians are also found to be a very closed ethnic group* by both indicators. Consequently, the results for Albanians and Roma do not facilitate the indicator selection. We also do not put much emphasis on the results for Bosniaks and Yugoslavians because their ethnic categories lack consistency over the 45-year period analyzed.

Table 5 about here

Now, we turn to interpreting the values in the upper part of Table 4. The *Coleman– Liu–Lu-indicator* suggests that the segmentation of the marriage market was the *strongest* between Serbs and non-Serbs (CLL = 0.63), followed by Croats and non-Croats (CLL = 0.59). The degree of segmentation is weaker for Czechs (CLL = 0.32), Hungarians (CLL = 0.36), Italians (CLL = 0.20), and Slovaks (CLL = 0.27). By comparison, the conventional log *odds-ratio* produces a less intuitive ranking. For instance, it suggests that ethnic endogamy is the *weakest* among Croats (log OR=3.17), while Serbs rank only slightly higher (with log OR=3.96).

Table 6 about here

The results obtained with the odds-ratio are difficult to reconcile with *survey evidence* on social distances shown by Tables 5 and 6. For our research, the most important aspect of social distance – among those assessed by the surveys – is the acceptance rate of marriage with members of various ethnicities. By focusing on this dimension, we find that (i) Croats in Croatia are typically more open to accept intermarriages with Italians than with Serbs, and (ii) Serbs in Serbia are typically more open to accept intermarriages with Hungarians, Romanians, and Slovaks than with Croats.

For example, while 58.4% of the respondents from the Croat majority in Croatia express willingness to accept an Italian to marry a close relative, only 40.6% would accept a Serb relative-by-law (see Table 5). Also, while 55.3%, 55.8%, and 61.1% of Serb respondents in Serbia express willingness to accept a Hungarian, Romanian, or Slovak marriage partner, respectively, only 54.9% would accept a Croat spouse (see Table 6).

Turning to *perceived community attitudes* in Serbia, we find that the pattern we identified is even more pronounced. In particular, only 21.7% of Serb respondents believe that the Croat community would accept marriage with a Serb, compared to 36.0% for Hungarian, 22.6% for Romanian, and 41.2% for Slovak community (see Table 6). Similarly, perceived acceptance of friendship and co-working relationships also shows that Hungarians, Romanians, and Slovaks are socially less distant from Serbs than Croats.

These survey evidence substantiate the finding by Gregurović (2018) that ethnic social distances measured in Croatia and Serbia are similar to each other. At the same time, they challenge the results by Mrdjen and Bahnik (2018) that ethnic endogamy in Croatia would be the weakest in relation to Croats and non-Croats (mostly Serbs), while being second weakest for Serbs and non-Serbs (mostly Croats). More importantly, the empirical findings support the methodological choice of relying on the Coleman–Liu–Lu-indicator: it not only has attractive analytical and empirical properties presented by Liu and Lu (2006) and Naszodi (2025) but also delivers a Croatia-specific ranking of ethnic endogamies that is consistent with survey-based social distance indicators.

# 5. Conclusion

In this paper, we have examined the segmentation of the marriage market in Croatia between 1970 and 2015, focusing on ethnic endogamy of various groups. Our analysis highlights that the choice of indicator substantially affects the relative ethnic endogamy identified. Moreover, the indicator-specific assessments of the ethnic divides are not equally plausible. While the conventional (log) odds-ratio produces an assessment that is counter-intuitive, the Coleman–Liu–Lu-indicator documents a pattern of ethnic segmentation of the Croatian marriage market that is consistent with survey evidence on the relative social distances between different ethnic groups. It is subject to future research to test whether the Coleman–Liu–Lu-indicator outperforms the odds-ratio in other settings as well, where both historical events and survey evidence offer empirical selection criteria for the suitable indicator.

## Tables

Table 1: Population by ethnicity according to censuses in Croatia, 1971–2011

| Nationality | Number of population | | | | | Percentage distribution (%) | | | | |
|---|---|---|---|---|---|---|---|---|---|---|
| | 1971 | 1981 | 1991 | 2001 | 2011 | 1971 | 1981 | 1991 | 2001 | 2011 |
| **Upper panel: focus ethnicities** | | | | | | | | | | |
| Czechs | 19,001 | 15,061 | 13,086 | 10,510 | 9,641 | 0.4 | 0.3 | 0.3 | 0.2 | 0.2 |
| Hungarians | 35,488 | 25,439 | 22,355 | 16,595 | 14,048 | 0.8 | 0.6 | 0.5 | 0.4 | 0.3 |
| Italians | 17,433 | 11,661 | 21,303 | 19,636 | 17,807 | 0.4 | 0.3 | 0.5 | 0.4 | 0.4 |
| Slovaks | 6,482 | 6,533 | 5,606 | 4,712 | 4,753 | 0.1 | 0.1 | 0.1 | 0.1 | 0.1 |
| Croats | 3,513,647 | 3,454,661 | 3,736,356 | 3,977,171 | 3,874,321 | 79.4 | 75.1 | 78.1 | 89.6 | 90.4 |
| Serbs | 626,789 | 531,502 | 581,663 | 201,631 | 186,633 | 14.2 | 11.6 | 12.2 | 4.5 | 4.4 |
| **Lower panel: ethnicities not analyzed** | | | | | | | | | | |
| Albanians[a] | 4,175 | 6,006 | 12,032 | 15,082 | 17,513 | 0.1 | 0.1 | 0.3 | 0.3 | 0.4 |
| Bosniaks[b] | 18,457 | 23,740 | 43,469 | 20,755 | 31,479 | 0.4 | 0.5 | 0.9 | 0.5 | 0.7 |
| Roma[a] | 1,257 | 3,858 | 6,695 | 9,463 | 16,975 | 0.0 | 0.1 | 0.1 | 0.2 | 0.4 |
| Yugoslavs[b] | 84,118 | 379,057 | 106,041 | - | - | 1.9 | 8.2 | 2.2 | - | - |
| Others[c] | 99,374 | 143,951 | 235,659 | 161,905 | 111,719 | 2.2 | 3.1 | 4.8 | 3.8 | 2.7 |
| **Total population** | 4,426,221 | 4,601,469 | 4,784,265 | 4,437,460 | 4,284,889 | 100.0 | 100.0 | 100.0 | 100.0 | 100.0 |

*Source:* Partial reproduction of Table 1 in Mrdjen and Bahnik (2018). Original source: Statistical Yearbook of the Republic of Croatia 2001; Census of Population, Households and Dwellings, 2011. Croatian Bureau of Statistics, Zagreb.
*Notes:* *a*: Indicates ethnicities that are found to be very closed by both the odds-ratio and Coleman–Liu–Lu-indicator. *b*: Indicates ethnicities for which data reliability is limited. *c*: Comprises other groups (Austrians, Bulgarians, Montenegrins, Macedonians, Germans, Poles, Romanians, Russians, Ruthenians, Slovenians, Turks, Ukrainians, Vlachs, Jews) and persons who declared regional affiliation, religion, ethnically uncommitted, and unknown ethnicity. Since 2001 Yugoslavs and Muslims have been included in this category.

Table 2: Proportion (%) of ethnically endogamous marriages in total number of marriages by ethnic groups and by gender in Croatia, 1970–2015

| Gender | Alb. | Bosn. | Croats | Czechs | Hung. | Ital. | Roma | Serbs | Slov. | Yugo. |
|---|---|---|---|---|---|---|---|---|---|---|
| Total | 40.5 | 31.1 | 86.8 | 17.8 | 20.4 | 9.8 | 70.6 | 49.6 | 16.8 | 52.7 |
| Men | 50.5 | 42.0 | 92.4 | 32.7 | 36.7 | 16.2 | 81.9 | 65.8 | 26.8 | 71.0 |
| Women | 67.2 | 52.5 | 93.5 | 28.1 | 31.5 | 20.1 | 83.6 | 66.9 | 22.4 | 67.1 |

*Source*: Partial reproduction of Table 2 in Mrdjen and Bahnik (2018). Original source of data: Demografska statistika (1970–1990), Table Marriages by nationality of women and brides, Federal Statistical Office, Belgrade, and on Data set 6-1-4 (1991–2015), Croatian Bureau of Statistics, Zagreb.

Table 3: Odds ratios computed for all ethnic groups in Croatia, 1970–2015

| Alb. | Bosn. | Croats | Czechs | Hung. | Ital. | Roma | Serbs | Slov. | Yugo. |
|---|---|---|---|---|---|---|---|---|---|
| 1,389.6 | 214.2 | 23.8 | 208.1 | 162.5 | 108.0 | 16,597.4 | 52.7 | 499.8 | 152.6 |

*Source*: Partial reproduction of Table 3 in Mrdjen and Bahnik (2018).

Table 4: Log odds-ratio and Coleman–Liu–Lu-indicator by ethnic groups in Croatia, 1970–2015

| | **Upper panel: focused ethnicities** | | | | | |
|---|---|---|---|---|---|---|
| **Measure** | Croats | Czechs | Hung. | Ital. | Serbs | Slov. |
| Log odds-ratio | 3.17 | 5.34 | 5.09 | 4.68 | 3.96 | 6.21 |
| Coleman–Liu-Lu-indicator | 0.59 | 0.32 | 0.36 | 0.20 | 0.63 | 0.27 |
| **Lower panel: ethnicities outside of our focus** | | | | | | |
| **Measure** | | | | | | |
| | Alb.[b] | Bosn.[a] | Roma.[b] | Yugo.[a] | | |
| Log odds-ratio | 7.24 | 5.37 | 9.72 | 5.03 | | |
| Coleman–Liu-Lu-indicator | 0.67 | 0.52 | 0.84 | 0.70 | | |

*Notes:* *a*: Indicates ethnicities for which data reliability is limited (see the first criterion discussed in the introduction). *b*: Indicates ethnicities that are found to be very closed by both the odds-ratio and the Coleman–Liu–Lu-indicator (see the second criterion discussed in the introduction).

Table 5: Percentage of Croat ethnic majority respondents in Croatia reporting that they would be willing to accept a member of a particular ethnic minority group in the stated social relations

| Minorities | Peer relations | | | | | Position of power | |
|---|---|---|---|---|---|---|---|
| | Living in country | Neighbor | Workmate | Friend | **Close family by marriage** | Boss | Political leader |
| Bosnian Muslims | 64.2 | 63.1 | 63.9 | 57.6 | **33.6** | 46.1 | 22.5 |
| Serbs | 67.3 | 64.6 | 63.7 | 59.1 | **40.6** | 45.8 | 22.3 |
| Italians | 78.0 | 78.3 | 78.4 | 75.9 | **58.4** | 62.3 | 31.3 |
| Roma | 62.3 | 54.9 | 57.1 | 50.2 | **27.3** | 41.3 | 21.1 |
| Albanians | 62.4 | 60.0 | 60.2 | 53.6 | **31.3** | 43.9 | 23.3 |
| Slovenes | 76.8 | 76.1 | 77.5 | 74.1 | **58.6** | 60.6 | 30.9 |

*Source:* Reproduction of Table 3.2 in Strabac and Valenta (2016). Original source: South-East European Social Survey Project (SEESP) – data were collected during the period November 2003 to March 2004.

*Notes:* Columns show percentages of Croat respondents who would accept a member of a given minority living in the country, neighbor, workmate, friend, close family by marriage; the last two columns report acceptance as boss and as political leader.

Table 6: Social distancing based on personal and perceived attitudes among Serbs (%).

| Type of relation | Albanians | | Bosniaks | | Croats | | Hungarians | | Roma | | Romanians | | Slovaks | |
|---|---|---|---|---|---|---|---|---|---|---|---|---|---|---|
| | PA | PCA | PA | PCA | PA | PCA | PA | PCA | PA | PCA | PA | PCA | PA | PCA |
| Cohabitation | 63.9 | 25.0 | 80.2 | 50.1 | 79.2 | 42.6 | 85.7 | 61.2 | 79.5 | 44.4 | 85.6 | 60.0 | 88.8 | 66.1 |
| Neighborhood | 62.5 | 21.7 | 80.0 | 48.6 | 79.5 | 41.8 | 86.2 | 60.7 | 65.5 | 25.1 | 85.5 | 59.2 | 88.8 | 65.3 |
| Acquaintances | 67.6 | 24.1 | 83.2 | 51.1 | 82.4 | 44.7 | 88.6 | 63.2 | 75.6 | 31.2 | 87.6 | 59.9 | 90.8 | 65.8 |
| Co-workers | 67.2 | 24.5 | 83.1 | 51.7 | 81.9 | 46.0 | 87.4 | 63.1 | 76.1 | 30.5 | 86.8 | 60.5 | 90.3 | 65.8 |
| Friends | 62.4 | 19.1 | 80.1 | 43.1 | 80.2 | 37.2 | 86.7 | 56.9 | 70.1 | 24.4 | 85.6 | 55.1 | 88.9 | 62.1 |
| **Marriage** | **31.1** | **6.8** | **46.5** | **19.7** | **54.9** | **21.7** | **55.3** | **36.0** | **28.5** | **7.0** | **55.8** | **22.6** | **61.1** | **41.2** |
| Ranking – marriage (PA) | 6 | | 5 | | 4 | | 3 | | 7 | | 2 | | 1 | |
| Ranking – marriage (PCA) | | 7 | | 5 | | 4 | | 2 | | 6 | | 3 | | 1 |

*Source:* Partial replication of the table presented by Chart 10 of Bašić et al. (2020).
*Notes:* PA: Personal attitudes. PCA: Perceived community attitudes. Values represent the share (%) of Serb respondents who *completely agree* with the statement. The survey was conducted in 2020.